\documentstyle[11pt,newpasp_s217,twoside,epsf]{article}
\def\lesssim{\mathrel{\hbox{\rlap{\hbox{\lower4pt\hbox{$\sim$}}}\hbox{$<$}}}}
\def\gtrsim{\mathrel{\hbox{\rlap{\hbox{\lower4pt\hbox{$\sim$}}}\hbox{$>$}}}}

\newcommand{\as}{\mbox{$''$}}
\newcommand{\etal}{{\it et~al.\/}}
\newcommand{\EWHa}{\mbox{$EW{\rm(H{\footnotesize \alpha})}$}}
\newcommand{\HI}{\mbox {\sc H i}}
\newcommand{\HII}{\mbox {\sc H ii}}
\newcommand{\Hline}[1]{\mbox{H{\footnotesize {#1}}}}
\newcommand{\Halpha}{\Hline{\mbox{$\alpha$}}}
\newcommand{\kms}{\mbox{km\thinspace s$^{-1}$}}
\newcommand{\MHI}{\mbox{${\cal M}_{\rm HI}$}}
\newcommand{\Msun}{\mbox{${\cal M}_\odot$}}
\newcommand{\SFRD}{\mbox{$\dot{\rho}_{\rm SFR}(0)$}} 
\newcommand{\SHa}{\mbox{$S_e(H{\footnotesize \alpha})$}}
\newcommand{\tgas}{\mbox{$\tau_{\rm gas}$}}

\def\edcomment#1{\iffalse\marginpar{\raggedright\sl#1\/}\else\relax\fi}
\reversemarginpar

\begin{document}
\title{Starbursts and Extra-planar H$\alpha$ from SINGG}
 \author{Gerhardt R.\ Meurer}
\affil{Department of Physics and Astronomy, The Johns Hopkins
 University, Baltimore, MD 21218}

\begin{abstract}
The NOAO Survey for Ionization in Neutral Gas Galaxies (SINGG) is the
largest star formation survey of an \HI\ selected sample.  Since the
selection is made without regard to optical morphology, it is not biased
toward or against ``interesting'' types of galaxies; thus SINGG is an
ideal sample for studying galaxy demographics.  Of a sample of 90
extra-galactic sources observed in photometric conditions, all are
detected in \Halpha.  This indicates that dormant galaxies, those
containing an appreciable ISM but no star formation, are at best rare.
We have made first pass morphological surveys for starbursts, as judged
by \Halpha\ surface brightness, and outflows as judged by extra-planar
\Halpha.  We find that about 15\%\ of the sources contain starbursts, with
little dependence on the neutral hydrogen mass \MHI.  Nearly one half of
a sample $\sim 35$ edge-on galaxies show evidence for extra-planar
\Halpha\ having a scale size of 0.5 Kpc or larger, while nearly one
quarter have extra-planar \Halpha\ features 1.0 Kpc in size or larger.
There is a hint that high \MHI\ systems preferentially have displaced
outflows (chimneys, or fountains) while central outflows (galactic
winds) preferentially occur in low \MHI\ systems.  However, a larger
sample (e.g.\ the full SINGG survey) is needed to confirm this trend.
\end{abstract}

\section{Introduction to SINGG}

Here I present preliminary morphological results from {\em SINGG\/} -
the Survey for Ionization in Neutral Gas Galaxies.  SINGG is an approved
and ongoing survey project of the National Optical Astronomy Observatoryical
(NOAO).  It is a follow-up survey to HIPASS - the \HI\ Parkes All Sky
Survey (Staveley-Smith \etal\ 1997).  HIPASS has surveyed the entire
southern sky for neutral hydrogen emission out to nearly 13000 \kms.
SINGG is intended to look for star formation, as traced by \Halpha\
emission, in this \HI\ selected sample.

The primary motivation for SINGG is to measure the star formation rate
density of the local universe, \SFRD.  The selection by 21cm emission
bypasses all the common {\em optical\/} selection biases (e.g.\ dust
content and surface brightness), which should allow a more complete and
representative survey of local star formation.  The planned
incorporation of ultraviolet data from {\em GALEX\/} and infrared data
from {\em IRAS\/} and {\em SIRTF\/} will allow a multi-wavelength
determination of \SFRD, addressing issues of dust extinction and the
initial mass function.  SINGG is also intended to examine the \HII\
region luminosity function, determining how it varies as a function of
morphology and \HI\ mass, \MHI.

In this paper I address some of the morphological questions that
motivate SINGG.  In particular (1) What fraction of gas rich galaxies
are dormant?  That is, what fraction contains no detectable star
formation?  (2) How common are starbursts?  And, (3) how common are
galactic winds?  Addressing these questions will give us a better
understanding of the life-cycle of star formation, and how star forming
galaxies enrich the inter-galactic medium.

After presenting a few more details of the SINGG survey in \S{2}, I
discuss the \Halpha\ detectability of HIPASS sources in \S{3}.  In \S{4}
I present plans for making an objectively defined starburst sample and
estimate the starburst incident rate. In \S{5} I summarize a
morphological survey for extra-planar \Halpha.  Finally \S{6} summarizes
these preliminary results.

\section{The SINGG survey: more details}

The SINGG sample is selected from the HIPASS survey
completely blind to the UV through infrared properties of the sources.
A total of 471 targets with a peak \HI\ flux density $f_\nu \ge 0.05$ Jy
in the Parkes 21cm spectrum were chosen by \MHI\ so as to have a flat
distribution in log(\MHI).  This is impossible at the lowest and
highest masses, where the HIPASS sample is volume limited.  Otherwise, we
have preferentially selected the nearest galaxies in each \MHI\ bin so
as to better resolve \HII\ regions.  Figure~1 shows the \MHI\
distribution of the final sample.

\begin{figure}[h]
\plotfiddle{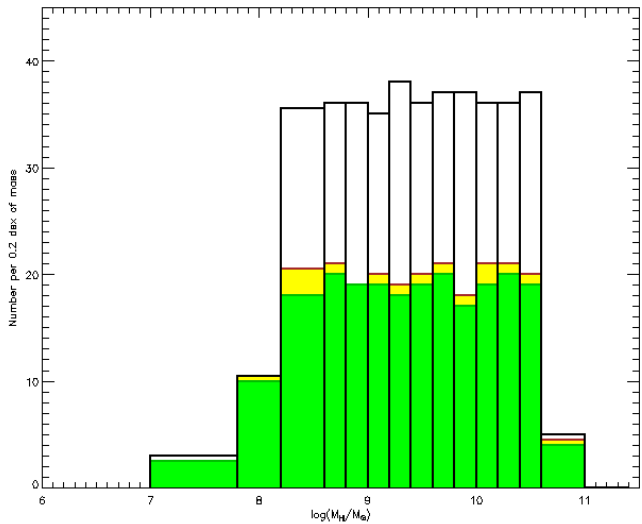}{7.0cm}{0.0}{40}{40}{-150}{0}
\caption{The \HI\ mass histogram of the final SINGG sample. Shading
indicates the portion observed in \Halpha\ as of July 2003.  Lighter
shading indicates sources that need further observations.}
\end{figure}

The observations with the CTIO 1.5m typically comprise three exposures
in a narrow band filter (matched to the wavelength of \Halpha\ at the
target's velocity) and a broad band $R$ filter, for total exposure times
of 30 and 6 minutes respectively.  We use specialized software from the
high-$z$ supernova group to align the images, combine the images within
each filter and subtract the scaled $R$-band image to produce the net
\Halpha\ image.  The final images typically have 1.5\as\ seeing.  The
$5\sigma$ point source detection limit is typically $F_{\rm H\alpha}
\sim 4 \times 10^{-16} \, {\rm erg\, cm^{-2}\, s^{-1}}$.  On scales
$\gtrsim 15''$ the images are typically flat to better than 1\%\ of the
sky level allowing large scale features with $S_{\rm H\alpha} \gtrsim 3
\times 10^{-18} \, {\rm erg\, cm^{-2}\, s^{-1}\, arcsec^{-2}}$ (0.6
Rayleigh) to be discerned at the $5\sigma$ level after sufficient
smoothing.

\section{Star formation in \HI\ selected galaxies}

We find that all extra-galactic HIPASS targets display \Halpha\
emission.  This result is based on data from four observing runs,
comprising 90 extra-galactic HIPASS targets (this excludes two frames
which targeted High Velocity Clouds and one frame strongly affected by
twilight sky).  The result is summarized in Fig.~2.  All HIPASS targets
are comprised of at least one \Halpha\ emitting galaxy, defined as a
spatially extended and distinct source in the \Halpha\ and/or $R$ image
having readily detected net \Halpha\ emission.  In all cases, much of
the \Halpha\ emission is organized into knots or high surface brightness
concentrations -- \HII\ regions or starbursts.  Hence, all targets are
undergoing some star formation.

\begin{figure}[h]
\plottwo{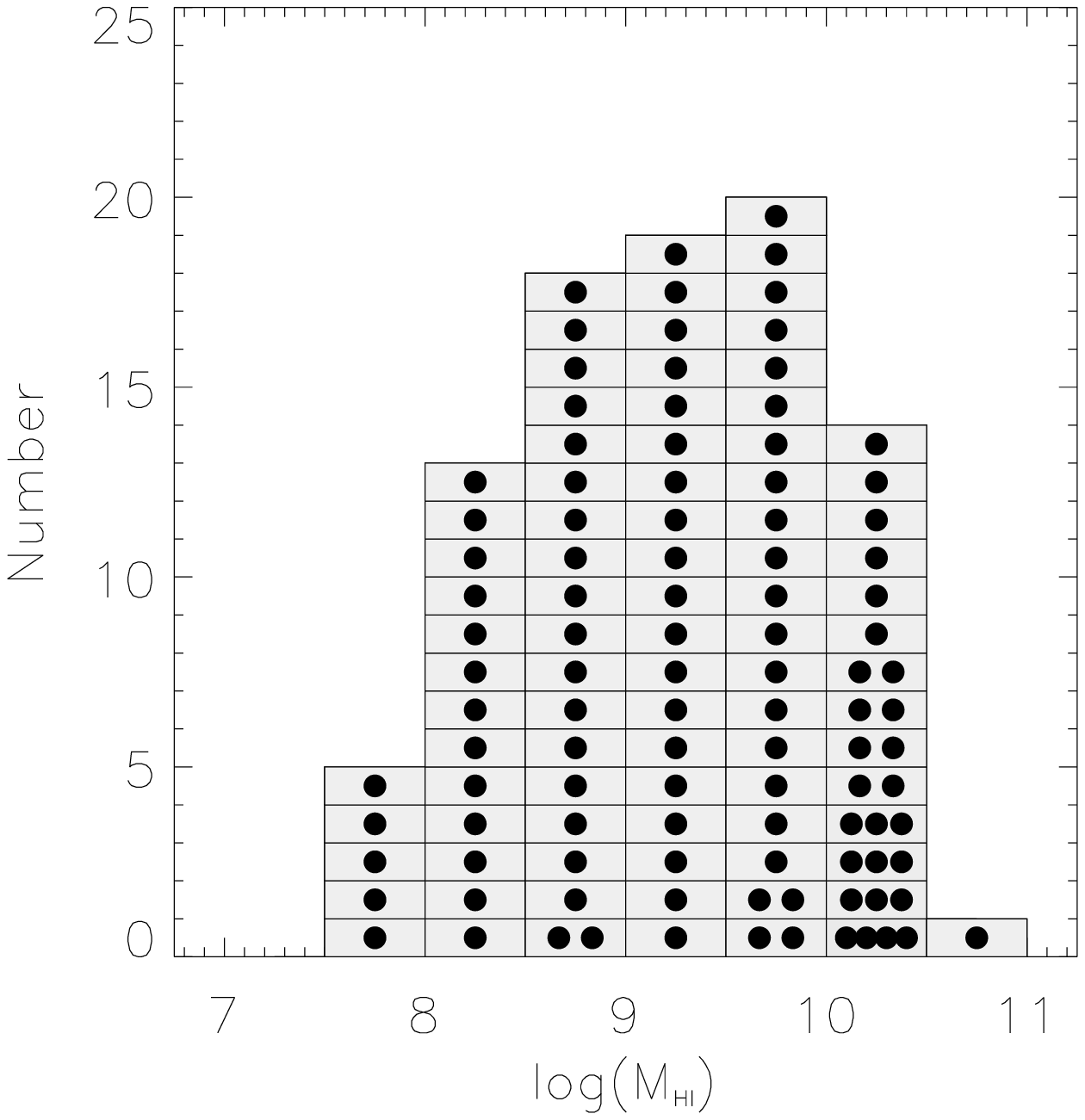}{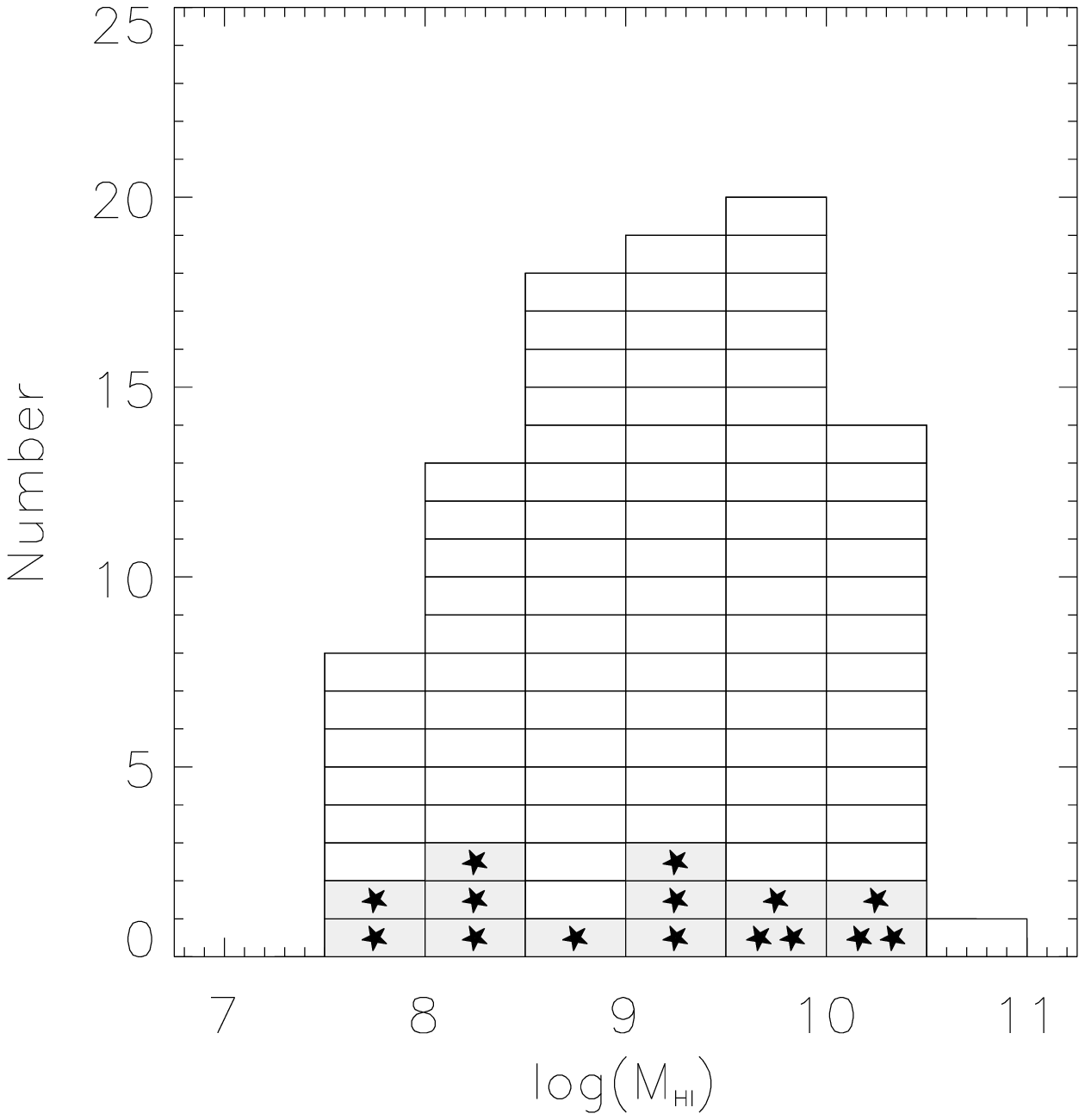}
\caption{a (left). The \HI\ mass histogram of 90 SINGG targets observed
over four observing runs.  Each rectangle represents one HIPASS target,
while each dot within a rectangle represents an \Halpha\ emitting
galaxy.  b (right), the same histogram, marking the targets containing
starburst galaxies.  Each starburst galaxy is indicated with a $\star$. }
\end{figure}

While extra-galactic \HI\ imaging surveys often turn up sources that were
previously uncataloged in optical surveys, free-floating \HI\ sources
with no optical counterpart are rare (Zwaan \etal\ 1997; Ryan-Weber
\etal\ 2002).  One possible exception is HIPASS J1712-64 (Kilborn \etal\
2000).  However, even in this case the possibility that this may be an
extreme version of an HVC bound to our galaxy has not been ruled out.
The dearth of free-floating extra-galactic \HI\ may be due to the fact
that when there is sufficient \HI\ for a gas cloud to be self
gravitating, it is gravitationally unstable until some
stars form so that they can heat the ISM enough to arrest further star
formation. Low mass \HI\ clouds that are not self-gravitating would have
low column density and are susceptible to ionization by the UV background
(Zwaan \etal\ 1997).  So \HI\ is either associated with stars or
destroyed.

Our results allow a stronger statement - dormant galaxies are rare or do
not exist.  That is, if a galaxy has an ISM with $\MHI \gtrsim 3\times
10^7 \Msun$ it also has {\em recently\/} (within 10 Myr) formed
high mass stars.  The gravitational instability in the ISM is not halted
by feedback from evolved stellar populations.  Instead new stars
continue to form, including the massive stars that ionize \HII\ regions.

While some star formation is always found, the range of star formation
morphologies wide.  We find low surface brightness dwarfs with a few
faint \HII\ regions, giant spirals with morphologies ranging from
flocculent to grand-design, high central surface brightness starbursts
(discussed below), and residual star formation in early type disk
galaxies.  SINGG also finds many multiple sources, particularly for
$\MHI > 10^{9.5} \Msun$ (Fig 2a.).  The higher multiplicity fraction,
may in part be due to the generally larger distance of the highest mass
sources and thus larger projected field.  In many cases the companions
are obvious, having similar angular extent and a merging, colliding or
interacting morphology.  In other cases the companions are less obvious
small sources, easily mistaken for field galaxies, save for their
\Halpha\ emission.  I conclude that narrow band imaging is an effective
means to locate companion galaxies.

The SINGG images also frequently reveal unresolved emission line sources
projected far from the primary \Halpha\ source on the frame.  The images
alone are incapable of discriminating what these emission line dots, or
``ELdots'' are.  Possibilities include outer disk or halo \HII\ regions
(Ferguson, \etal\ 1998), very faint companion galaxies (e.g.\ similar to
those proposed by Blitz \etal\ 1999), or background emission line
galaxies (Boroson, Salzer \&\ Trotter 1993).  In her contribution, Emma
Ryan-Weber presents results from our first spectroscopic follow-up
observations of ELdots.

\section{Starbursts with SINGG\/}

Starburst galaxies are systems which are experiencing a short duration
episode of intense star formation.  Beyond this statement the term
starburst is not really well defined in the literature.  The SINGG
sample selection has netted many well known galaxies which have been
called starbursts.  These include M83, NGC~1705, NGC~1808, and NGC~5253.
All have high central surface brightness, and frequently have a
galactic wind morphology.

One goal of SINGG is to provide an objectively defined starburst sample.
We will look at three quantities often used to define starbursts:
\Halpha\ equivalent width, \EWHa, effective \Halpha\ surface brightness,
\SHa, and gas consumption timescale, \tgas.  A high \EWHa\ indicates a
large SFR compared to the past average. High values of \SHa\ indicates
currently intense star formation.  A short \tgas\ indicates that star
formation can not continue for long at its present rate.  Hence the
three measures of star formation are complementary.  Our plan is to use
the well known starbursts in the sample as a training set to determine
the best cuts in these quantities for our objectively defined starburst
sample.

Currently we have defined a sample of starbursts by \SHa.  The incidence
of these starbursts as a function \MHI\ is given in Fig.~2b.  The sample
includes well known starbursts, and sources selected to have similar
surface brightness.  This yields a a first pass starburst incidence rate
of $13/90 \sim 15$\% of SINGG targets containing at least one starburst
galaxy.

\section{Extra-planar \Halpha}

To address the incidence rate of extra-planar \Halpha, or outflows, we
have compiled a database of the largest possibly expanding structures in
each galaxy.  These have morphologies including bubbles, rays or
chimneys, and edge-brightened and filled fans with or without caps.  The
database notes whether the source is in the center (i.e.\ a galactic
wind) or outer regions (e.g.\ a fountain or chimney) of the galaxy, the
source's radius $r$ and orientation relative to the host's minor axis,
its morphology and finally whether the source is seen in \Halpha\
emission (almost all cases) or continuum absorption (just a few cases).
In order to focus on extra-planar ISM we limited ourselves to
structures with a minor axis extension, and with hosts having axial
ratio $a/b > 2$.  We considered two samples, structures having $r > 0.5$
Kpc, and those with $r > 1.0$ Kpc.  We limit ourselves to sources with
$r > 5''$ so that the structures are well resolved, hence the two
samples are distance limited to within 21 and 42 Mpc respectively ($H_0
= 70\, {\rm km\, s^{-1}\, Mpc^{-1}}$).

\begin{figure}[h]
\plottwo{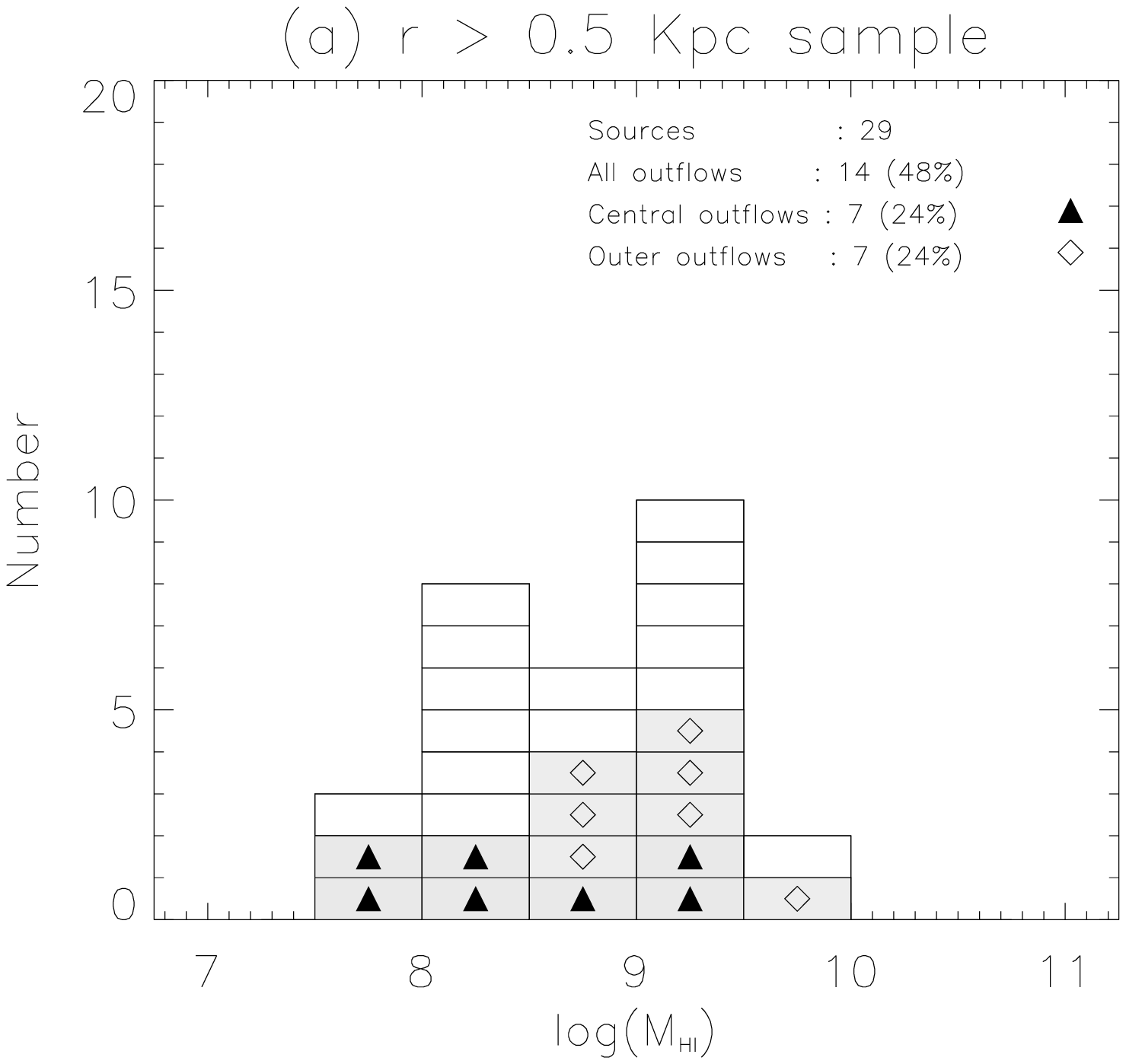}{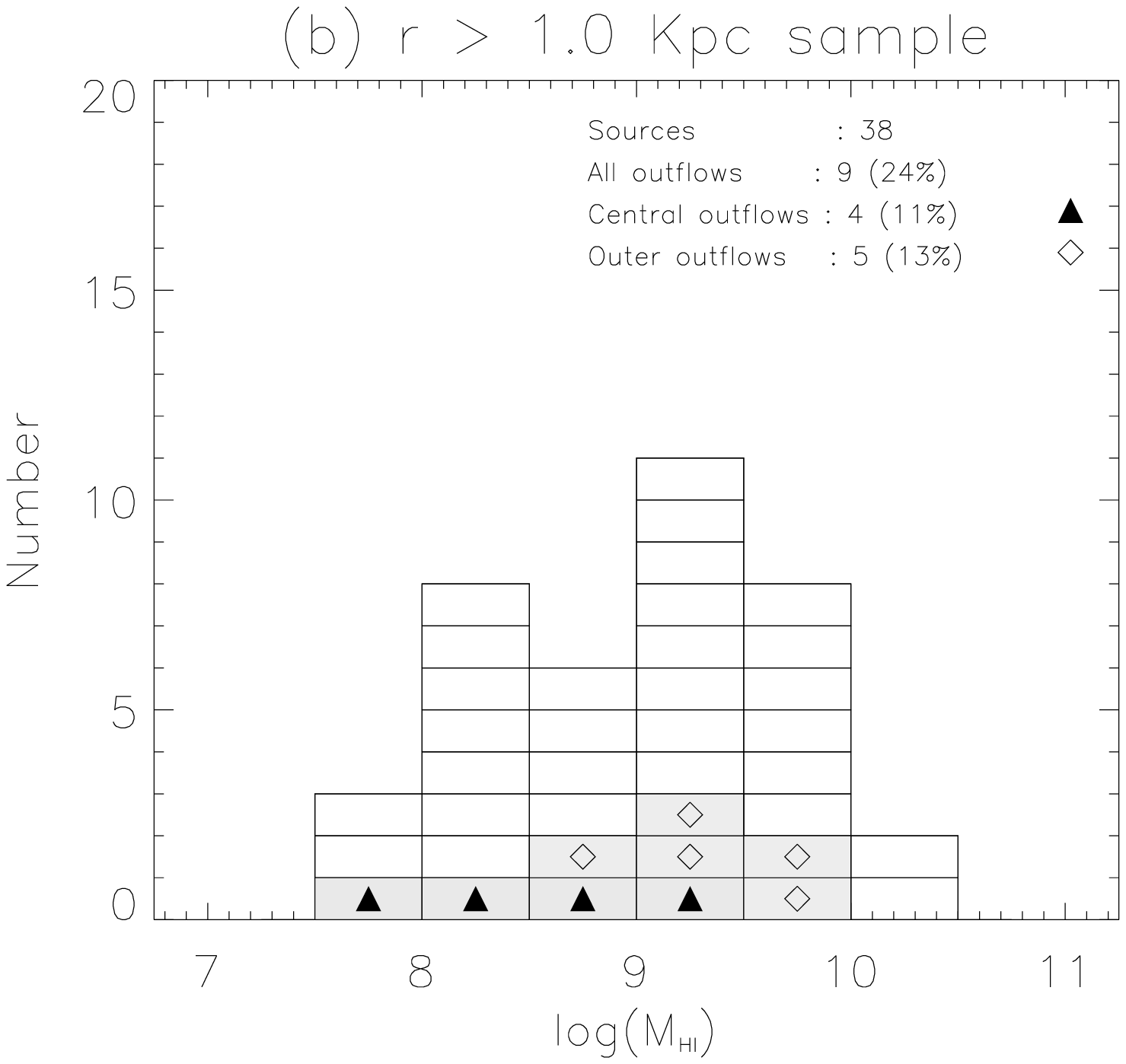}
\caption{Mass histograms showing the incidence of central and outer
outflows having scale sizes (a) $r > 0.5$ Kpc, and (b) $r > 1.0$
Kpc. Central outflows are marked with filled triangles while outer
outflows are indicated with diamonds.  }
\end{figure}

Figure 3a and 3b show \MHI\ histograms of the \Halpha\ emitting galaxies
meeting the distance and $a/b$ cuts for the two samples.  Sources are
distinguished by whether the largest expanding structure is central or
in the outer regions of the galaxy.  Nearly half of the
hosts have extra-planar structures with $r > 0.5$ kpc, while about a
quarter have structures larger than 1.0 Kpc.  The galaxies with
extra-planar \Halpha\ are not all starbursts.  This is true even if we
limit ourselves to just the cases with central outflows, which are also
found in galaxies with modest values of \SHa.

The histograms imply that the central outflows tend to be in lower \MHI\
hosts, while larger mass systems tend to host fountains.  However, we
have found high \HI\ mass \Halpha\ emitting galaxies that have strong
galactic winds that just don't quite make our distance and $a/b$ cuts.
One case is the well known central starburst in NGC~1808
($\log(\MHI/\Msun) = 9.4$) which is one of the two cases where the
outflow is best traced by dust lanes.  However, its $a/b \approx
1.5$, is too round to make our selection.  The three galaxies with the
largest galactic winds ($r = 6$ to 12 Kpc) seen in the database are all
contained within a single Hickson Compact Group (HCG~16, combined
$\log(\MHI/\Msun) = 10.3$).  However, with $D = 54$ Mpc, it does not
make our distance cut.  These examples show that there are indeed high
mass systems with central galactic winds, however we need a larger
sample to find them in statistically significant quantities.

\section{Summary}

The strategy of SINGG is to survey for star formation in the galaxies
known to contain its essential ingredient, interstellar hydrogen, i.e.\ we
know where the fuel is; that's where to look for the fire.  This has
proven to be a very effective approach, since we find that there are no
dormant galaxies: when a galaxy has $\MHI > 3\times 10^7 \, \Msun$ it
is observed to be undergoing some star formation.  The amount of \HI\
does not prove to be a good predictor of star formation morphology which
varies greatly at fixed \MHI.  About 15\%\ of \HI\ selected
galaxies are starbursts.  Extra-planar \Halpha\ also appears to be a
common phenomenon with half of edge-on galaxies having features with a
radius $\geq 0.5$ Kpc, and about a quarter having a radius $\geq 1$
Kpc.  

\acknowledgments{I thank the other members of the SINGG collaboration
for their contribution to the survey. They are Harry Ferguson, Rachel
Webster, Michael Frinkwater, Ken Freeman, Dan Hanish, Tim Heckman, Rob
Kennicutt, Virginia Kilborn, Patricia Knezek, B\"arbel Koribalski,
Martin Meyer, Sally Oey, Mary Putman, Emma Ryan-Weber, Chris Smith,
Lister Staveley-Smith and Martin Zwaan.  I am particularly grateful to
Dan Hanish for doing much of the data reduction and to Patricia Knezek
to reading this manuscript and suggesting improvements.}

\end{document}